# A Framework for Scalable Digital Twin Deployment in Smart Campus Building Facility Management

Thyda Siv

Graduate Student, School of Building Construction, Georgia Institute of Technology.
Email: thyda@gatech.edu

## ABSTRACT

Digital twin (DT) offers possibilities for enhancing facility management (FM) in campus settings. However, existing research often focuses narrowly on single domains—such as point-cloud geometry or energy analytics, without providing a scalable, interoperable workflow that integrates building geometry, equipment metadata, and operational data into a unified FM platform. This study proposes a comprehensive framework for scalable digital-twin deployment in smart campus buildings by integrating 3D laser scanning, BIM modeling, and IoT-enabled data visualization to support FM operations and maintenance. The methodology includes (1) reality capture using terrestrial laser scanning and structured point-cloud processing; (2) development of an enriched BIM model incorporating architectural, mechanical, electrical, plumbing, conveying, and sensor systems; and (3) creation of a DT environment that links equipment metadata, maintenance policies, and simulated IoT data within a digital-twin management platform. A case study of the Price Gilbert Building at Georgia Tech demonstrates the implementation of this workflow. A total of 509 equipment items were modeled and embedded with Omniclass classifications into DT. Ten interactive dashboards were developed to visualize system performance. Results show that the proposed framework enables centralized asset documentation, improved visibility of building systems, and enhanced preventive and reactive maintenance workflows. Although most IoT data were simulated due to limited existing sensors, the prototype validates the feasibility of a scalable DT for FM and establishes a reference model of DT for real-time monitoring, analytics integration, and future autonomous building operations.

# 1. INTRODUCTION

Facility Management is related to operation, maintenance and optimization of building and infrastructure. According to the International Facility Management Association (IFMA), facility management is the process that involved with people, place, processes, and technology [1]. Traditional FM practices often rely on manual inspections, old-paper-based documentats, and reactive maintenance, which can cause fragmentation in the practice of FM [2]. In the context of large campus buildings, facility managers face several challenges such as outdated equipment, and inconsistent maintenance, which significantly affect operations and user satisfaction [3]. In recent years, there has been rapid advancement of digital technologies such as Building Information Modeling (BIM), the Internet of Things (IoT), and digital twin (DT) [4]. With this technology available, the way modern buildings are managed, operated and maintained has been transformed in a more efficient and sustainable way.

In campus facility management, digital twin technology offers transformative capabilities that enable educational institutions to transition towards smarter, more responsive, and sustainable campus environments [5]. By linking real-time sensor data to 3D digital models, facility managers can visualize building performance dynamically, enabling early detection of anomalies and preventive maintenance before equipment failures occur [6]. Additionally, the integration supports cost tracking and energy optimization by providing data-driven insights into energy usage and maintenance expenses that improve overall efficiency and flatten demand levels [7]. The integration with BIM enables facility managers to prioritize tasks, quickly identify system issues within the virtual model, and make informed decisions that extend asset life, reduce waste, and promote sustainability [8]. Digital twins also enhance efficiency by reducing reliance on paper-based documentation such as maintenance records, sensor data, and asset histories, to be centrally stored and easily accessed over the building's lifecycle [9].

This paper proposes a framework for digital twin creation for facility management in campus settings. The specific objectives of this study are: (1) To propose a framework for creating a digital twin of campus buildings for facility management through the integration of laser scanning, BIM, and IoT sensors. (2) To examine how visualization with real-time data of crital building system equipment and historical building documentation within a digital twin platform can enable sustainable building operations and maintenance in campus facility management.

The rest of this paper is organized as follows: Section 2 provides a comprehensive review of the current practices in facilities management and the challenges of facilities management for campus, followed by an introduction to the relevant concepts of digital twin and why its application is beneficial in addressing the aforementioned problems. Section 3 outlines the research methodology employed in this study. Section 4 presents the prototype of the digital twin. Section 5 presents the discussion of the findings, followed by limitations of the research, and future research directions are discussed in Section 6. Finally, the conclusion is provided in Section 7.

## 2. LITERATURE REVIEW

### 2.1 FACILITY MANAGEMENT AND ITS CHALLENGES IN CAMPUS SETTINGS

Facility management (FM) in campus environments plays an important role by ensuring the efficient operational and sustainability of academic buildings. Universities typically manage large and diverse building portfolios, such as classrooms, libraries, laboratories, residence halls, dining spaces, athletic facilities, and mechanical plants. Each of these buildings has a distinct operational demands. This diversity, combined with high occupant turnover and continuous building usage, places significant pressure on FM teams.

Traditional practices in campus FM rely heavily on manual inspections, paper-based documentation, and reactive maintenance workflows. These approaches create several persistent challenges. First, facility managers lack real-time visibility into building performance, imposing challenges in dynamic system monitoring and early anomaly detection that could prevent equipment failures[10]. Second, reactive maintenance leads to unplanned downtime and higher long-term expenses[11]. Third, fragmented asset information, with visual asset data, maintenance records, and system documentation scattered across paper-based and digital systems, makes it challenging for maintenance workers to access critical information in a useful form[12]. Fourth, traditional methods provide limited capability for energy monitoring, resulting in rising electricity costs and inefficient usage patterns[13]. These limitations underscore the need for advanced technologies—such as digital twins—that provide real-time monitoring, predictive maintenance capabilities, and integrated data management to address these challenges.

### 2.2 DIGITAL TWIN IN CAMPUS BUILDINGS

Digital twin (DT) technologies have gained growing attention as a means to enhance building performance monitoring, operational decision-making, and asset management[14]. A digital twin functions as a dynamic virtual representation of a physical building, updated through the integration of BIM models and IoT metadata[15]. In campus environments, DTs integrate fragmented building data, enable real-time environmental and system monitoring, support predictive maintenance workflows, and enhance energy efficiency to advance sustainability objectives[16]. Lu et al. has shown the potential of DTs at both the building and city scale by integrating BIM, GIS, and IoT data streams into a unified digital environment, illustrating how multi-layered digital twin platforms can enhance urban services, spatial analytics, and real-time facility operations in complex built environments.

Recent advancements in digital twin technology have demonstrated diverse approaches to integrating real-time data, building information, and predictive analytics for facility management applications. Digital twin enablers, including application tools, data storage and communication networks, are critical for FM applications such as real-time operations monitoring, energy management, and predictive maintenance[18]. A point-cloud–driven workflow is presented to generate as-is building data and support more accurate DT/BIM representation and FM space analytics[19]. In addition, a tool-based DT architecture is employed, showing how multi-sensor streaming through Power BI, Azure IoT, and custom web interfaces can be used to support rapid diagnostics, as demonstrated in an operating-room case[20]. The study integrates IoT data, fiber-optic sensing, and BIM with an LSTM prediction model to enable energy forecasting on a smart campus, achieving strong prediction accuracy and measurable reductions in energy consumption[17]. Also, the study shows that combining digital twins with mobile IoT sensing and UWB indoor positioning to compute and visualize spatially dynamic thermal comfort conditions in real time[21].

However, there remain significant gaps in the literature, particularly the absence of reference models at the building scale that clearly define, how existing buildings should be captured, modeled, and transformed into a functional digital twin. In addition, AI has been used within digital twin frameworks to support key FM functions, such as anomaly detection, energy prediction, spatial analytics, and predictive maintenance. However, existing studies rarely specify how these capabilities apply to essential building systems. In particular, prior work does not clearly detail how AI-driven DTs can monitor, or optimize core MEP equipment such as HVAC components, electrical distribution systems, air-handling units, lighting controls, or plumbing fixtures.

To address these gaps, this study develops a scalable digital twin creation framework that integrates 3D point cloud data from laser scanning, BIM modeling, and interactive dashboard visualization within a unified DT platform. The framework encompasses mechanical, electrical, plumbing, conveying, and occupancy systems to enable comprehensive facility management of building operation. The framework also demonstrates how FM workflows, including asset documentation, preventive maintenance and reactive maintenance, can be embedded directly into a single digital twin platform. In short, this study links traditional as-built documentation with precise geometric modeling and real-time operational data to support more informed, data-driven O&M decision-making in campus buildings.

## 3. METHODOLOGY

### 3.1 LASER SCANNING

This study employs 3D laser scanning technology to capture accurate spatial data of the existing campus building, then generate a point cloud file that will be used for 3D model development.
A terrestrial laser scanner (TLS), such as the Leica BLK360, FARO Focus S350, or Trimble X7, was used to capture the existing interior of the building. The laser scanning process involves 3 steps: (1) acquisition, (2) registration, and (3) processing of point cloud.

(1) Prior to scanning, as-built drawings were reviewed, and a site visit was conducted to verify whether the existing conditions matched the drawings. After this, a scanning plan was developed, which included the planned scanner positions and target arrangements. Targets such as checkerboards or spheres were strategically placed to assist in the later alignment between scans. The scanner was positioned at multiple locations, each capturing a 360° panoramic sweep of its surroundings. The three most commonly adjusted parameters during scanning were resolution, quality, and color capture. Each scan required a different duration depending on these parameter settings. A total of N scans were performed to ensure complete site coverage. All raw scan data were stored in the proprietary scanner format files (*.fls*, *.ptx*, or *.e57*).

(2) During registration, all scans were imported into the registration software (Leica Cyclone or FARO Scene). Using either target-based or cloud-to-cloud registration, the scans were aligned into a single coordinate system. Next, noise and irrelevant data (people, or moving objects) were removed using filtering algorithms or manual editing. The cleaned dataset was then cropped to retain only the region of interest. The data were then merged to form a single unified point cloud file (.rcp, .e57, or .las). The final point cloud were achieved, with an accurate representation of the as-built geometry of the campus building.

(3) The final point cloud was imported into Autodesk ReCap for visual inspection, which then converted into a BIM model. Accuracy assessments were conducted by comparing selected dimensions in the point cloud with physical site measurements. All processed files were exported in standard formats (.e57, .rcs, .las, or .ply) to ensure interoperability with BIM and GIS platforms. Metadata such as scanner type, date, coordinate system, and registration error were documented, and data were stored securely on institutional servers.

## 3.2 BIM MODEL DEVELOPMENT

Following the laser scanning and data processing, this phase focuses on the development of a three-dimensional architectural model using the captured point cloud data. The processed and registered point cloud was saved as a ReCap project file (.RCP) which then import to Autodesk Revit. A new Revit project was created with an appropriate project template and units corresponding to the site's coordinate system. Each floor plan level was established according to the elevation reference extracted from the scan data. Revit's Auto-Origin to Origin or By Shared Coordinates method was applied to ensure alignment between the model and the point cloud coordinate system. Architectural elements such as walls, doors, windows, floors, ceilings, and roofs were modeled by tracing the point cloud in plan and elevation views to accurately represent the as-built conditions. For areas that were not captured during scanning, 2D floor plans and field measurements were referenced to reconstruct missing spaces and ensure model completeness.

Additional building systems and IoT sensors were integrated within the Revit environment. Building systems such as HVAC, lighting, and indoor air quality monitoring—were modeled and annotated with unique identifiers and sensor attributes.

In addition to the geometric modeling, relevant data and metadata were incorporated to transform the 3D representation into a BIM model suitable for facility management (FM). The inclusion of these data layers enabled connectivity between the digital environment and real-time physical data streams. This stage elevated the model from a static BIM to the next maturity level of the digital twin (DT). The finalized Revit model was then saved and exported in interoperable formats (RVT, IFC, and NWC) which serve as the 3D representation of the digital twin.

## 3.3. DIGITAL TWIN DEVELOPMENT

After the architectural Revit model was created and enriched with metadata, building systems, and sensor information, the next stage involved developing the Digital Twin (DT) environment. The DT serves as a dynamic representation of the physical building, integrating both the BIM model and real-time operational data from IoT systems.

*First*, the Revit model, containing architectural, structural, and MEP systems, was exported to an interoperable format such as IFC or gbXML to enable connection with external data management platforms. Each system and component was linked to metadata, including equipment IDs, maintenance schedules, and sensor types. A network of low-cost IoT sensors was deployed across representative spaces within the building. For instance, $CO_2$ sensors – DHT22 (AM2302) or BME280; Carbon dioxide monitoring – Sensirion SCD30 or MH-Z19C; temperature and humidity sensors – PIR HC-SR501, was chosen based on accuracy, protocol, and environmental tolerance. Each sensor node was connected to a microcontroller (ESP32 or Raspberry Pi) that collected readings at 1–5 minute intervals. Data was transmitted wirelessly through Wi-Fi or MQTT protocol to a local IoT gateway or cloud service (ThingSpeak, Node-RED, or AWS IoT Core). The gateway served as the intermediary between the physical environment and the virtual twin.

*Second*, a digital twin management platform (Autodesk Tandem, Azure Digital Twins, or Wiretwin) can be used to host the BIM model and manage live data streams. A database or cloud service was configured to collect sensor data through APIs or MQTT protocols, ensuring that real-time data such as temperature, humidity, and $CO_2$ concentration could be synchronized with the virtual model.

*Third*, within the DT platform, each Revit element was assigned a unique identifier (GUID) that matched its corresponding entry in the sensor data repository. This enabled two-way data binding, allowing real-time performance values to be visualized directly within the 3D model. Dashboards were created to represent sensor states and environmental parameters that provide an understanding of critical building system performance.

## 4. PROTOTYPE OF DIGITAL TWIN

### 4.1 LASER SCANNING

This study uses Price Gilbert building at Georgia Tech's campus as a case study due to its dynamic occupancy patterns. The building includes four floors, a basement level, and a roof. The as-built drawings were provided, and floor plans of the building will be used to complement the scans.

The model of the FARO equipment used by Georgia Institute of Technology (Georgia Tech) is the FARO Cobalt Array Imager. Thus, the scanning plan was developed according to the scanner's operating range, field-of-view requirements, and target placement recommendations.

For each planned scanner location, draw 30 ft circles around each scan location, to ensure proper range and target coverage. Scanner positions should be defined so that every position collects as much data as possible, and so the number of required scans is reduced to a minimum. Make sure all areas and points are covered, and prover overlap between scans is provided. Target placement should include the type of target (checkerboard or sphere) and the expected height at which the target will be placed. Scan positions must be numbered in the order they will be performed. During the scanning process, checkerboard targets and spheres were placed at the locations defined in the plan, and the scanner was mounted on the tripod and initialized. A new project was created for each floor, and scans were performed in order, with spherical targets moved as needed once they were no longer required for registration in previous positions. All scan files (.fsl) were saved and transferred for subsequent data processing.

Next, in Faro Scene, create new projects that correspond to each floor. Import and process the raw scans in Faro Scene and register them using the automatic registration tool and target-based registration. If additional refinement is required, manual or visual registration techniques may also be applied following the guidelines provided in the Faro Scene training manual. After achieving a complete and accurate registration, prepare the required visualizations by using clipping boxes to clean the view, remove noise, and highlight relevant interior or exterior spaces. The unified point cloud is then exported following the naming format for each floor. Next, create an Autodesk account if needed and download Autodesk Recap. Create a new Recap project, import the registered scans, index them, and save the project. Finally, export the complete Recap project as an *.rcp* file for each level. This exported point cloud will be used in the subsequent phases of the BIM modelling workflow.

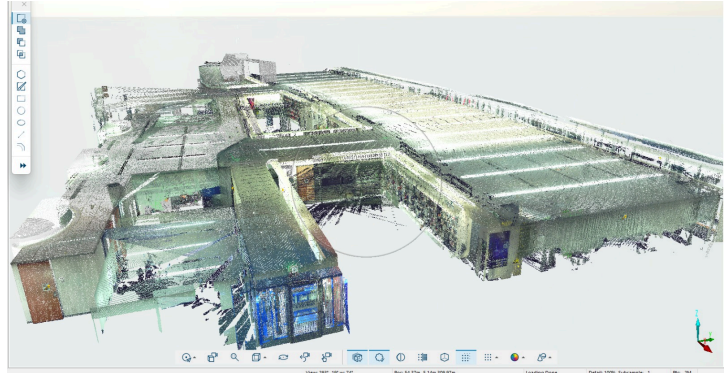
*Figure 1 The Unified Point Cloud (Autodesk ReCap)*

**4.2 BIM MODEL DEVELOPMENT**

In this stage, an architectural Revit model of the scanned floor is developed. The exported Recap file (.rcp) is imported into Revit as a point cloud reference, serving as the primary source for modelling the as-is geometry. Areas that were not directly captured during scanning (inaccessible rooms or occluded spaces) are modeled using existing floorplans to ensure spatial completeness. This combination of point cloud data and 2D drawings allows for the creation of an accurate and comprehensive architectural representation suitable for facility management applications.

To construct the Revit model, begin by creating a new project and importing the *.rcp* point cloud for each floor. Use the point cloud as the primary reference to accurately model all major architectural elements, including walls, floors, ceilings, doors, windows, stairs, storefront components, and any relevant built-in furniture. *Walls*: created by tracing point cloud surfaces in plan and elevation views. *Doors and Windows*: inserted based on visible openings in the point cloud; dimensions verified using point measurements. *Curtain walls or glazing systems*: modeled using components when applicable. *Floor slab*: generated by tracing the boundary visible in the point cloud. *Ceiling and roof:* modeled using section or elevation views to align with point cloud data. Height adjustments were made to match elevation points extracted from vertical scans. *Exterior conditions* may be approximated using onsite visual observations. Unscanned *interior areas* should be referred to the provided floorplans, and all *interior partitions* are represented with standard gypsum board finishes.

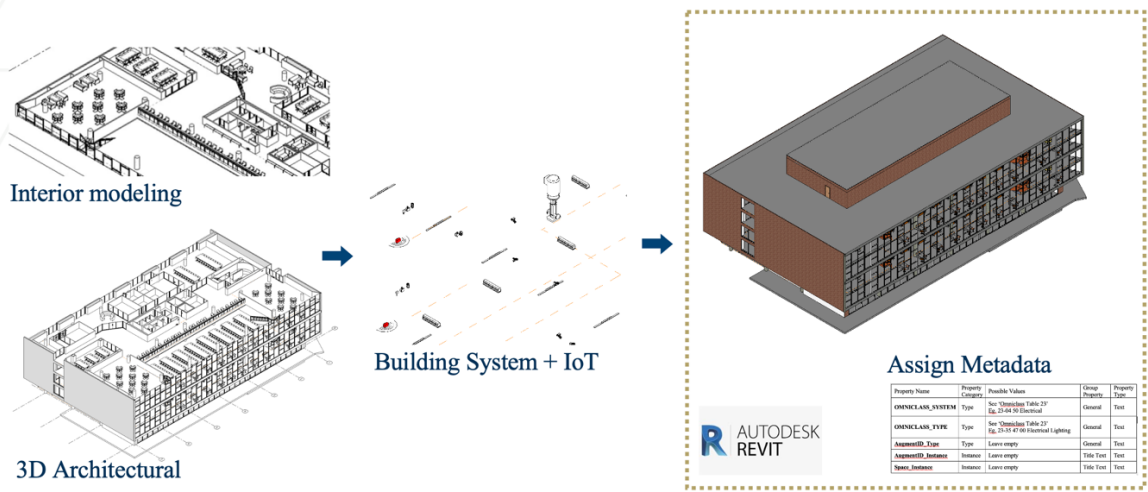
*Figure 2 BIM Model Development*

After completing the architectural modeling of the floor, the next step is to define and populate the room data within the Revit environment. Each room must be assigned Project Properties, and four instance properties are required for every room. (1) *Room-Category,* must be entered exactly as named and uses values from *Omniclass Table 15* to classify the room according to its functional use. For example, *13-55 11 00 Office Spaces*, *13-23 17 00 Restroom*, and *13-57 17 13 Break Room*. (2) *Room-Name,* contains a descriptive label for the room; while the name does not need to be unique, it should clearly communicate the room's purpose, such as *Men's RRs* or *Main Study Area*. (3*) Room-Tag,* contain a *unique value* for each room. For example, *Room 101*, *Room 230*, *140*, or *Restroom A*. (4) *Room-AugmentID* property is added but left blank. All four properties, Group Property under the *'General'* category and Property type under *'Text'*.

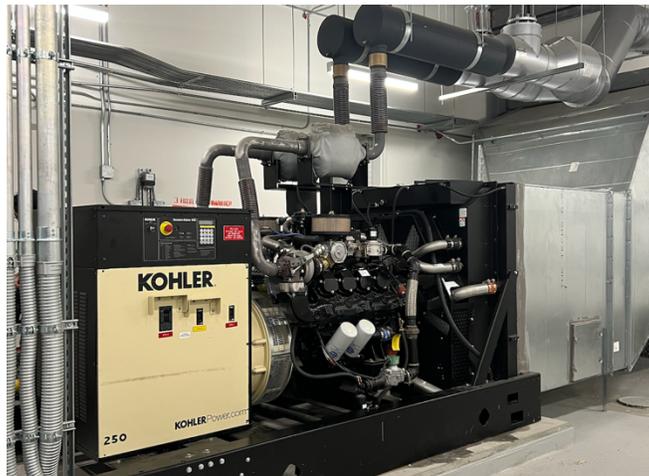

*Figure 3 Image of the generator captured during the building walkthrough*

Next, critical MEP components were incorporated into the model, and detailed metadata were assigned to each component. A walkthrough of the facility was conducted with the maintenance team to gain familiarity with building systems. The equipment incorporated into the BIM model represents key Mechanical, Electrical, Plumbing, Conveying, and sensor systems essential for facility management and future digital-twin integration. The following equipment were munually modeled:
    a) Mechanical: AHU, ERU, VAV box / VAV unit, hot water pumps, temperature sensors, humidity sensors, and carbon monoxide sensors.
    b) Electrical: lighting fixtures, generator, transformer, and distribution panel board.
    c) Plumbing: rainwater harvesting tank, preheat water tank, water pressure booster system, faucets, sinks, toilets, urinals, service sinks, water heaters, drinking fountains.
    d) Conveying: elevators
    e) Communication: occupancy sensors

Based on the equipment inventory, the above items were added into the Revit model, and several key properties must be included to ensure the equipment can be integrated into subsequent digital twin development (WireTwin). Each equipment must contain the following property names:

| Property Name | Property Category | Possible Values | Group Property | Property Type |
|---|---|---|---|---|
| **OMNICLASS_SYSTEM** | Type | See 'Omniclass Table 23'<br>Eg. 23-04 50 Electrical | General | Text |
| **OMNICLASS_TYPE** | Type | See 'Omniclass Table 23'<br>Eg. 23-35 47 00 Electrical Lighting | General | Text |
| **AugmentID_Type** | Type | Leave empty | General | Text |
| **AugmentID_Instance** | Instance | Leave empty | Title Text | Text |
| **Space_Instance** | Instance | Leave empty | Title Text | Text |

### 4.3 DIGITAL TWIN DEVELOPMENT

In this phase, WireTwin is used as the central platform for implementing the Digital Twin (DT). To create a Digital Twin for facility management, relevant O&M properties are added to each equipment type or instance, associated documents are linked, maintenance policies are created, reactive maintenance tasks are simulated, technician workflows are modeled, and visualization links are integrated to display simulated BAS-related equipment information.

In WireTwin, manufacturer documents such as cut sheets, operation manuals, warranty documents, and product specifications were uploaded using the *Equipment Module* to the corresponding equipment previously modeled in Revit. Next, define operational and maintenance properties for each equipment type or instance. These properties may include manufacturer name, equipment phases, water or energy consumption, capacity ratings, and dimensions. Populate these properties directly within the Equipment Module and use WireTwin's automatic extraction tool to populate some property fields based on uploaded documents. Using standards such as GSA maintenance guidelines or the manuals collected earlier, a maintenance policy for each equipment type, which describe the tasks, required frequency, start dates, and necessary resources or replacement parts. In the *Automations Module,* assign each maintenance policy to its corresponding equipment type. In addition, create a room-cleaning policies and assign them to different rooms.

Within the *Maintenance Module*, create reactive maintenance jobs associated with equipment and reactive jobs associated with building spaces. Each job should include a detailed description of the issue, the appropriate assignment (technician for equipment jobs and custodian for space jobs), and any required resources. Job details must align with the equipment type or space where the issue would realistically occur. Simulate the maintenance workflow by moving equipment-related reactive jobs and space-related reactive jobs through different status stages, such as ongoing, completed, or verified. Some automatically generated jobs created from maintenance policies should also be transitioned through stages. Add comments that represent typical technician or custodian actions based on the work performed.

Extend the Digital Twin by simulating BAS capabilities. For each item, add a sensor. Then, link each sensor property to a graphic or dashboard that visualizes real-time or simulated performance data, and configure the property to trigger an alarm when values fall outside expected operating parameters. To support enhanced facility-management insights, generate a broader set of maintenance jobs spanning multiple dates. Using the *Reporting Module*, produce analytics and graphical reports summarizing system performance, maintenance outcomes, equipment health, and staff activity.

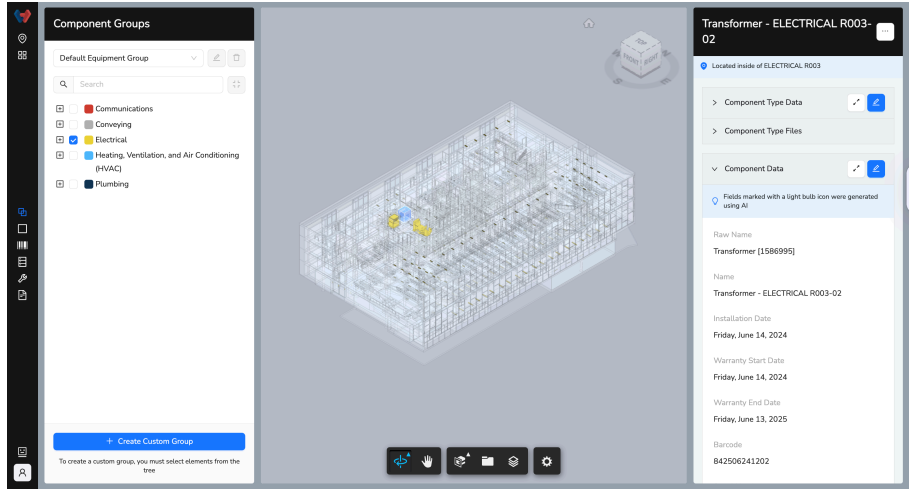

*Figure 4. Documentation integrated to Digital Twin Model on WireTwin*

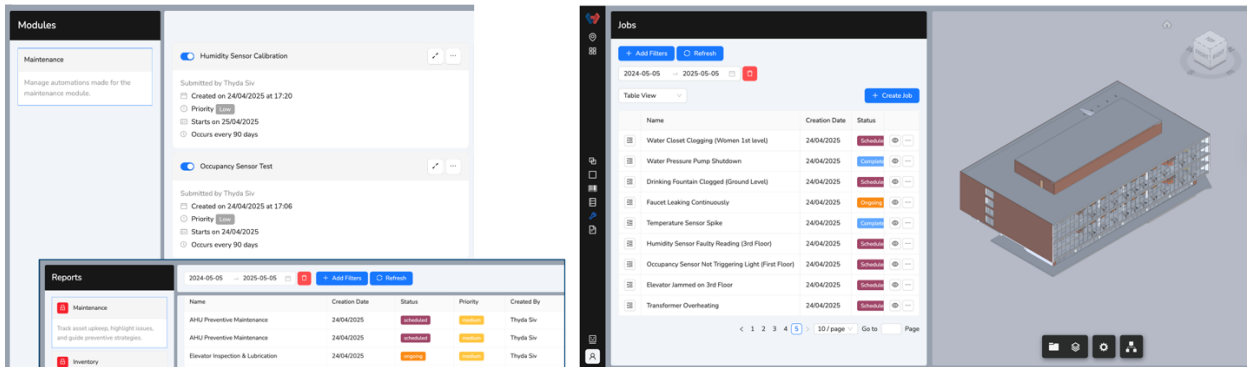

*Figure 5. O&M intergraed to Digital Twin Model on WireTwin*

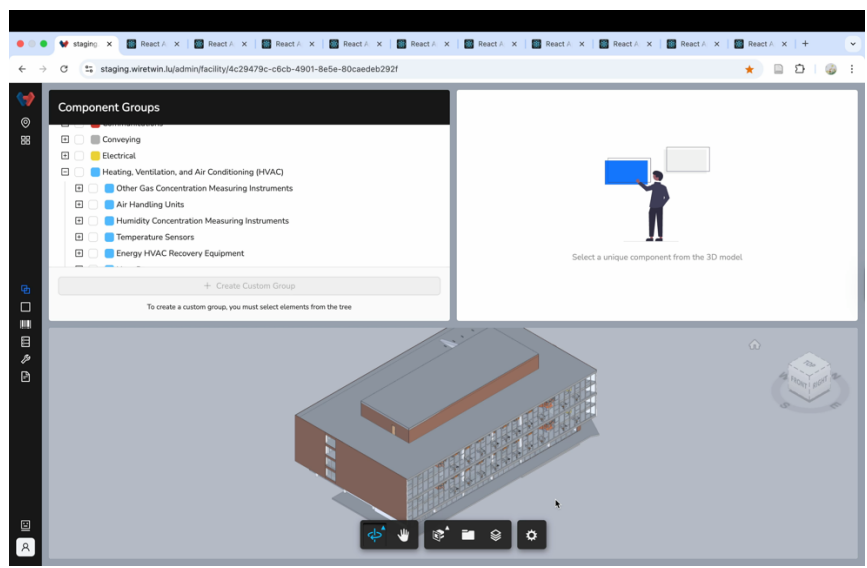

**Click here to view 'Demo of DT Dashboards'**

## 5. RESULT AND DISCUSSION

The prototype digital twin of the Price Gilbert building successfully integrated the 3D point cloud model, BIM environment, and facility-management data to create a functional representation of the building's systems. A total of *509 equipment items* were modeled across mechanical, electrical, plumbing, and conveying systems. This included 3 AHUs, 1 ERU, 1 VAV unit, 2 hot water pumps, 30 temperature sensors, 20 humidity sensors, 20 CO sensors, 300 lighting fixtures, 2 transformers, 16 faucets, 16 sinks, 16 toilets, 8 urinals, 4 service sinks, 8 water heaters, 2 drinking fountains, 2 elevators, 1 generator, and 60 occupancy sensors. Each item was assigned OMNICLASS codes, Augment IDs, and space-instance identifiers to enable interoperability and data tracking within the digital twin platform.

A major outcome of the implementation was the development of *10 interactive dashboards* representing the major building systems. These included:
1) *AHU Dashboard* – Airflow, supply/return temperature, humidity, and fault alerts.
2) *Drinking Fountain Dashboard* – Fixture usage patterns, filter condition, and maintenance cycles.
3) *Electrical Distribution Panel Board Dashboard* – Load distribution, energy consumption, and fault detection.
4) *Elevator Dashboard* – Operational status, trip logs, and maintenance history.
5) *Generator Dashboard* – Runtime hours, load capacity, fuel levels, and scheduled service.
6) *Lighting Dashboard* – Energy usage, on/off status, and occupancy-based control behavior.
7) *Temperature Sensor Dashboard* – Room-level temperature readings with deviation alerts.
8) *Transformer Dashboard* – Voltage, amperage, and power-conversion performance.
9) *Water Closet Dashboard* – Fixture condition, water-flow anomalies, and plumbing-related maintenance.
10) *Water Pressure System Dashboard* – Pressure stability, pump performance, flow rate, and system alerts.

It is important to note that most dashboard values and system behaviors were simulated, as the building did not have physical IoT sensors installed for HVAC, electrical, plumbing, or environmental monitoring at the time of this prototype. The only real sensor data available in the Price Gilbert building were from occupancy sensors, which were installed as part of the building's existing infrastructure. Therefore, dashboards related to HVAC performance, electrical loads, water pressure, plumbing fixtures, and environmental readings (temperature, humidity, and $CO_2$) were populated using simulated data to represent expected maintenance and operational conditions. The simulation allowed for evaluating the digital twin's visualization, equipment mapping, and FM workflow integration.

Despite these simulated inputs, the dashboards provided a clear representation of how real-time building operations could be monitored within a 3D digital environment. The system allowed equipment behavior to be visualized through color-coded overlays, alerts, and performance indicators, demonstrating the potential value of integrating actual sensor networks in future phases.

The prototype validated the feasibility of combining BIM models, metadata, and simulated operational data into a cohesive digital-twin platform suitable for facility management. The results show that digital twin technology can enhance preventive maintenance planning, improve equipment visibility, centralize building information, and support more informed decision-making in campus settings.

## 6. LIMITATIONS AND FUTURE RESEARCH

### 6.1 LIMITATIONS

Several limitations were identified during the development and implementation of the digital twin prototype. *First,* all values displayed on the digital-twin dashboards were simulated. Implementing a fully operational DT would require extensive work to procure, install, calibrate, and network physical HVAC, electrical, plumbing, and environmental sensors. These practical requirements demand significant time, cost, coordination with campus operations, and long-term maintenance considerations. Although the simulated dashboards effectively illustrate the intended real-time monitoring and analytics capabilities, the absence of live IoT data limited the ability to validate actual system behavior or evaluate automated FM alerts.

*Second*, the process of acquiring the 3D point cloud required a significant amount of time due to the need for developing a detailed scanning plan. The FARO Cobalt Array Imager required careful planning of scan positions, target placement, and sequencing to ensure adequate coverage and registration accuracy. This multi-step workflow, including target placement, scan overlap planning, and floor-by-floor alignment, extended the duration of the data collection phase. Newer generations of laser scanners, such as *FARO Orbis*, *Leica RTC360*, and *Trimble X12*, incorporate SLAM-based (simultaneous localization and mapping) technology and automated registration features that eliminate the need for manual target placement and significantly reduce planning time. Using such scanners in future work could accelerate the point-cloud acquisition process and minimize operational complexity.

*Third*, the prototype was developed using WireTwin as the primary digital-twin platform. While WireTwin provides flexibility for equipment-based modeling, maintenance workflows, and simulated sensor properties, it has limitations in terms of industry adoption, interoperability with existing BIM workflows, and long-term FM integration. Platforms such as Autodesk Tandem may provide additional benefits, including: *Native compatibility with Revit and IFC models*, enabling smoother updates and synchronization; *Direct asset mapping to Autodesk Construction Cloud workflows*, which enhances interoperability for large institutions; and *Improved support for real-time data connectors*, allowing future integration with actual IoT hubs and building management systems (BMS). Evaluating multiple platforms is necessary to determine the most scalable and industry-aligned solution for campus-wide deployment.

### 6.2 FUTURE RESEARCH

Integrating a wider range of physical IoT sensors, including temperature, humidity, $CO_2$ concentration, water pressure, vibration sensors for mechanical equipment, and energy meters, would enable real-time monitoring and allow a direct comparison between sensor readings and dashboard outputs. This would help assess the accuracy of live data visualization and determine whether predictive alerts and FM notifications align with actual building conditions.

The dashboards developed for this prototype can be extended beyond web-based visualization. Future implementation could focus on mobile applications or augmented reality (AR) interfaces, allowing technicians to access equipment status, alerts, and maintenance instructions while physically navigating the building. Mobile- and AR-enabled digital twins could significantly enhance response times and support hands-free FM operations.

More extensive user studies should be conducted with facility managers, maintenance teams, and campus administrators. Their feedback can guide improvements in interface design, data presentation, dashboard priorities, and integration requirements with existing FM workflows or CMMS systems.

## 7. CONCLUSION

This paper introduced a scalable framework for developing a digital twin for campus facility management by integrating 3D point cloud data, BIM modeling, and IoT-enabled dashboards. The framework outlines a systematic process beginning with laser scanning and point-cloud acquisition, followed by BIM enrichment using building systems and equipment metadata, and concluding with digital-twin implementation through a cloud-based platform. The approach demonstrates how campus buildings can transition from static documentation to dynamic, data-informed environments that support informed decision-making and long-term asset management.

The results of the study show that the proposed framework effectively captures building geometry, organizes equipment information, and visualizes system performance within an interactive digital-twin environment. A total of 509 equipment items were modeled and integrated into a unified BIM dataset, while ten dashboards were developed to represent key building systems, including HVAC, lighting, electrical distribution, and water systems. Although most dashboard data were simulated, the prototype successfully demonstrated how real-time information could be displayed, analyzed, and used to support preventive and reactive maintenance workflows.

The prototype validates the feasibility of using digital twins as a tool for enhancing campus facility management. It illustrates the potential benefits of combining BIM data, equipment metadata, and operational simulation into a single virtual environment. Despite limitations related to scanning duration, simulated sensor data, and platform constraints, the study provides an important foundation for future development. As more physical sensors are integrated and platform capabilities expand, the digital twin can evolve into a fully operational system that supports real-time monitoring, mobile accessibility, and predictive maintenance, ultimately contributing to more efficient, sustainable, and data-driven facility management practices.